# Launching Insights: A Pilot Study on Leveraging Real-World Observational Data from the Mayo Clinic Platform to Advance Clinical Research

Yue Yu, PhD[1], Xinyue Hu, MSc[1], Sivaraman Rajaganapathy, PhD[1], Jingna Feng, MSc[1], Ahmed Abdelhameed, PhD[1], Xiaodi Li, PhD[1], Jianfu Li, PhD[1], Ken Liu, MD, PhD[2], Liu Yang, MBBS[3], Nilufer Taner, MD, PhD[4], Phil Fiero, MSc[5], Soulmaz Boroumand, PhD[5], Richard Larsen, MSc[5], Maneesh Goyal, MS, MBA[5], Clark C. Otley, MD[5,6], Nansu Zong, PhD[1], John D. Halamka, MD[5*], Cui Tao, PhD[1,5*]

1. Department of Artificial Intelligence and Informatics, Mayo Clinic

2. MCHS Cardiology, Mayo Clinic

3. Division of Hepatology and Liver Transplant, Mayo Clinic

4. Department of Neurology, Mayo Clinic

5. Mayo Clinic Platform

6. Department of Dermatology, Mayo Clinic

## Abstract

**Backgrounds:** Artificial intelligence (AI) is transforming healthcare, yet translating AI models from theoretical frameworks to real-world clinical applications remains challenging. The Mayo Clinic Platform (MCP) was established to address these challenges by providing a scalable ecosystem that integrates real-world multiple modalities data from multiple institutions, advanced analytical tools, and secure computing environments to support clinical research and AI development. **Methods:** In this study, we conducted four research projects leveraging MCP's data infrastructure and analytical capabilities to demonstrate its potential in facilitating real-world evidence generation and AI-driven clinical insights. Utilizing MCP's tools and environment, we facilitated efficient cohort identification, data extraction, and subsequent statistical or AI-powered analyses. **Results:** The results underscore MCP's role in accelerating translational research by offering de-identified, standardized real-world data and facilitating AI model validation across diverse healthcare settings. Compared to Mayo's internal Electronic Health Record (EHR) data, MCP provides broader accessibility, enhanced data standardization, and multi-institutional integration, making it a valuable resource for both internal and external researchers. **Conclusion**: Looking ahead, MCP is well-positioned to transform clinical research through its scalable ecosystem, effectively bridging the divide between AI innovation and clinical deployment.

* Corresponding author: John D. Halamka, MD and Cui Tao, PhD

Future investigations will build upon this foundation, further exploring MCP's capacity to advance precision medicine and enhance patient outcomes.

**Introduction**

In recent years, artificial intelligence (AI) has emerged as a transformative force poised to revolutionize the field of biomedicine. However, the transition of AI algorithms from in silico simulations to practical, real-world clinical applications presents significant challenges. Effective implementation of AI in healthcare requires a comprehensive consideration of the entire ecosystem, extending beyond the algorithms themselves[1]. For instance, within the medical domain, a prominent trend is the development of multimodal AI models that integrate diverse data types across multiple modalities[2-4]. This advancement, however, introduces complex issues, such as safeguarding patient privacy amidst the aggregation of sensitive information[5]. From the perspective of AI model development, advancing beyond retrospective design and validation poses an additional hurdle[6,7]. Moreover, ensuring the accessibility of advanced tools and adequate computational resources to accommodate the varied requirements of users is essential for widespread adoption and effectiveness[1,8]. Another significant challenge is the integration of expert-in-the-loop systems that require no-code AI platforms[9], which are crucial for enabling non-technical medical professionals to effectively use and interact with AI tools without needing extensive programming knowledge.

To establish a comprehensive solution for expediting the development of medical AI, Mayo Clinic established the Mayo Clinic Platform (MCP)[10], which focuses on transforming healthcare through data science and digital health technologies. By leveraging a vast array of clinical data, advanced analytics, and collaborative networks like the Mayo Clinic Care Network, the platform aims to improve patient care and streamline health outcomes. It fosters innovation by enabling healthcare organizations, providers, and digital health companies to access real-time insights and deploy cutting-edge solutions.

In this study, we explore the platform's capabilities by conducting four research projects using real-world EHR data and tools from the MCP Discover. By leveraging MCP's robust data infrastructure and versatile tools, ranging from intuitive visualizers to AI workspaces, our projects have effectively demonstrated how the MCP environment can facilitate innovative research. These projects span from population-based analyses to the validation and development of advanced AI models, illustrating MCP's pivotal role in clinical decision support, optimizing clinical trials, and advancing translational medicine. The findings from these projects not only reinforce the critical importance of real-world data in clinical research but also underscore the transformative potential of AI applications in enhancing healthcare outcomes.

**Method**

**Real-World Observational Data in MCP**

MCP provides access to extensive, high-quality clinical data, including standardized structured data (e.g., diagnoses, lab results, medications) and unstructured data (e.g., clinical notes, images) via MCP Discover. This de-identified, multisite data spans diverse demographics and captures patient journeys over time. Currently, MCP Discover's datasets include over 13.6 million patient records, 3.9 billion images, 2.7 billion lab results, 12.9 million pathology reports, and 1.25 billion clinical notes from Mayo Clinic and Mercy, all accessible through a secure data science environment.

**MCP Tools used in this study**

MCP provides various tools to accommodate different needs. In this study, since we only used structured EHR data within MCP, the following tools were utilized.

**Cohort Visualizer** facilitates the quick creation, characterization, and comparison of patient cohorts for hypothesis testing and analysis using EHR data. It supports both structured and unstructured data, offering code-free analytics and intuitive visualization tools. Users can load or create new cohorts and analyze them using graphical or tabular formats by the cohort builder. With user-friendly navigation, it allows users, regardless of technical expertise, to explore vast clinical datasets using standard clinical codes or keywords, helping to accelerate clinical research and address unmet needs in translational medicine. Additionally, for more detailed downstream analysis, it provides SQL code to facilitate data retrieval from the EHR database. Figure 1 shows the user interface of MCP cohort builder.

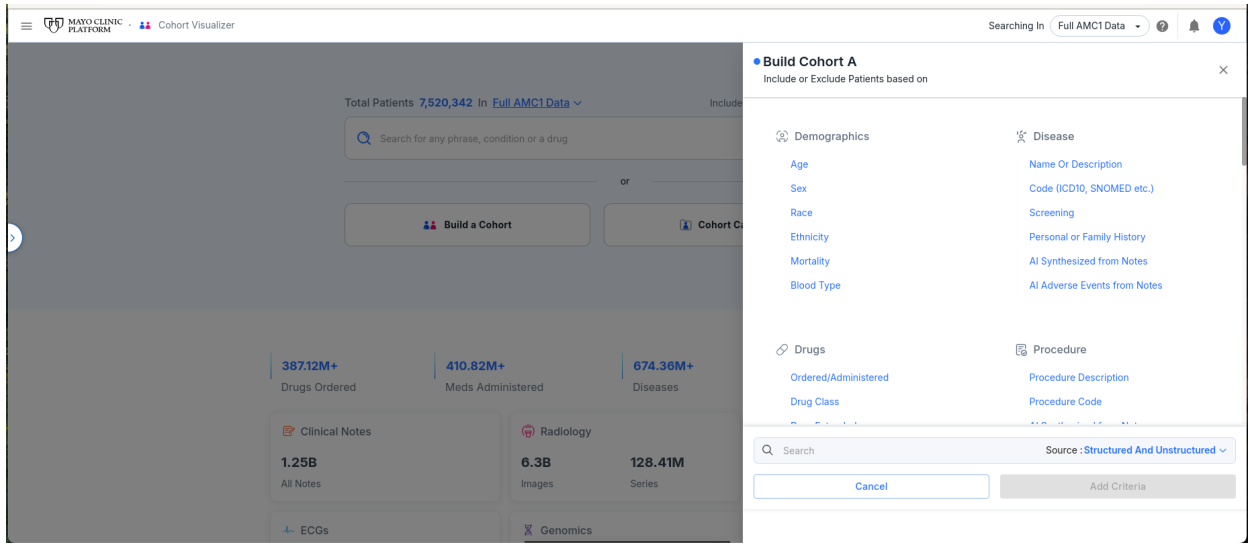

**Figure 1.** MCP Cohort builder interface.

**Schema Visualizer** provides an interactive interface for exploring the data dictionary and schema within MCP. It offers detailed information on tables, columns, and their relationships, along with query code examples for downstream data collection (Figure 2). Additionally, it features an advanced search tool that enables users to efficiently locate specific tables, columns, or values within the data schema.

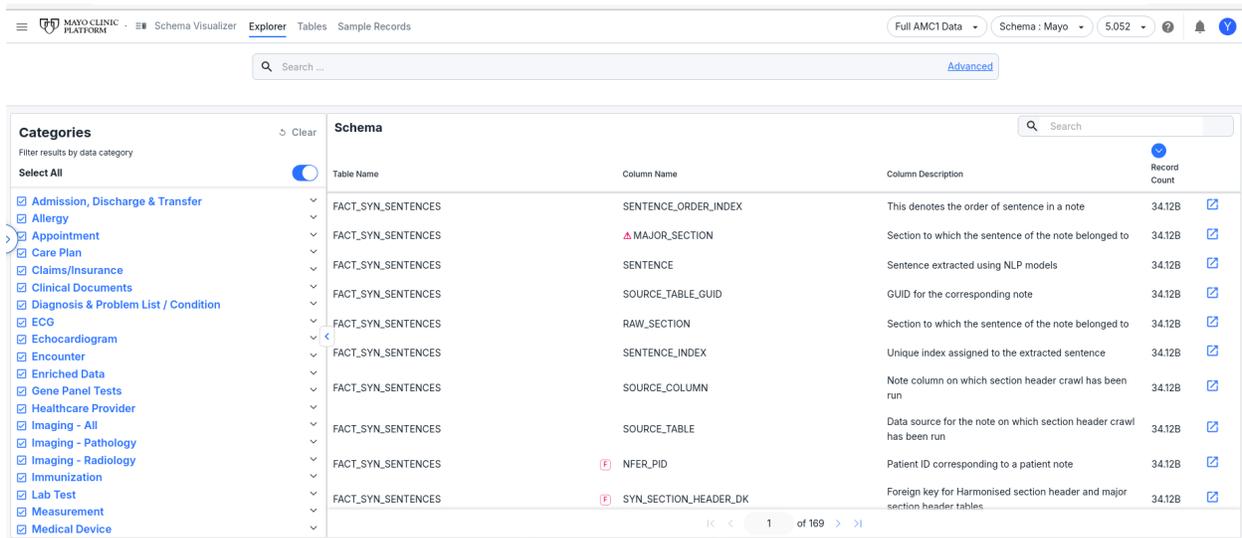

**Figure 2.** MCP Schema Visualizer interface.

**Workspaces** in MCP offer a comprehensive environment for accessing data and computing resources, supporting advanced analytics and data science workflows. They provide the latest open-source tools, packages, and libraries for cloud-based computation, with integrated support for JupyterLab and RStudio to accommodate diverse coding needs. This all-in-one platform streamlines data collection, processing, and analysis. Additionally, Workspaces include high-performance computing capabilities for resource-intensive tasks such as data mining, machine learning, and deep learning. They also offer code-level guidance for various applications, including data extraction, large language model (LLM) execution, and medical image processing. Furthermore, users can leverage Git within Workspaces to efficiently manage and collaborate on their repositories in GitHub. Figure 3 shows the launcher page of MCP workspace.

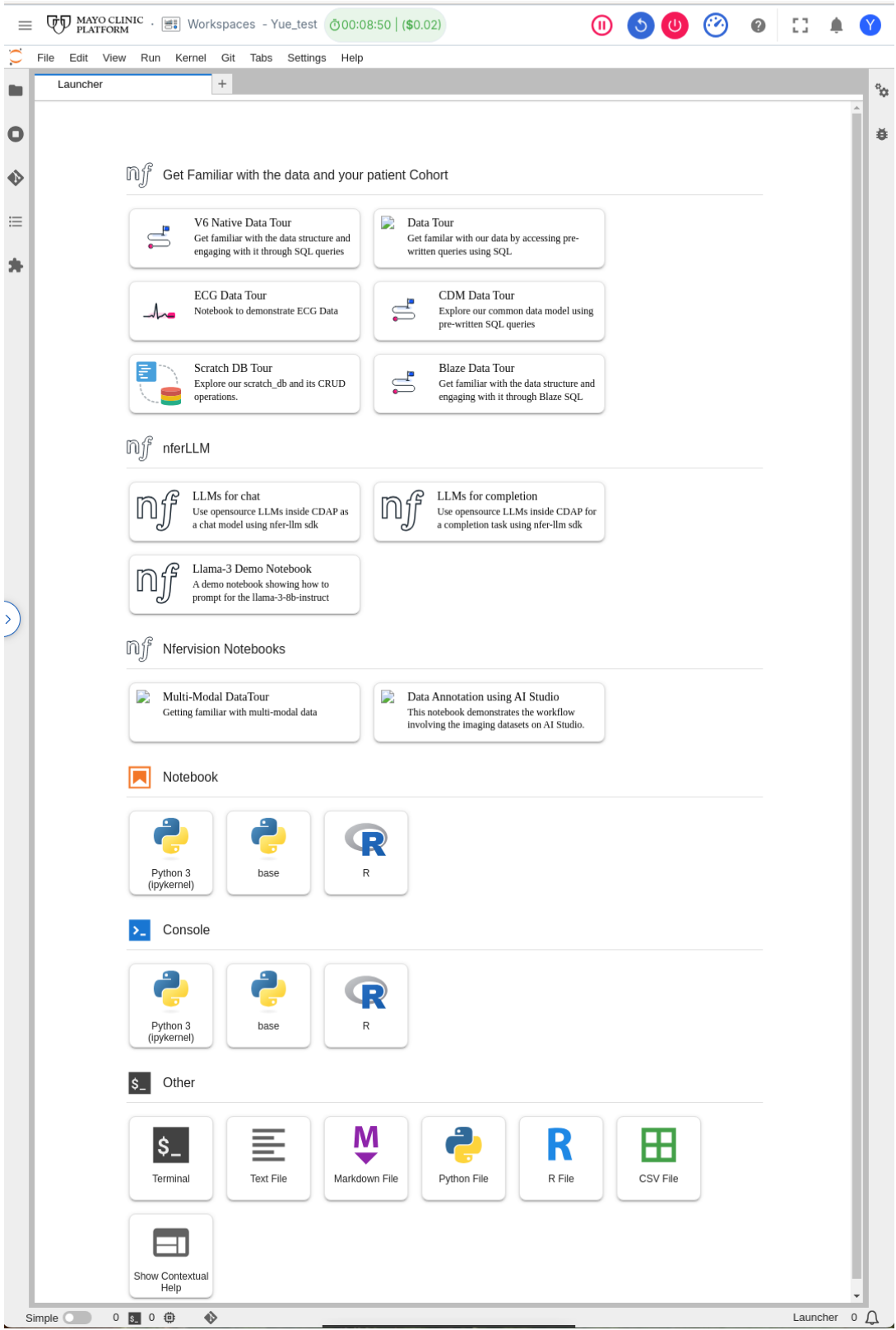

**Figure 3.** The launcher page of MCP Workspace.

**Research Projects Conducted on MCP**

**Project 1. Emulating drug efficacy randomized controlled trials (RCTs) for heart failure (HF) patients using real-world observational clinical data.** This project leverages the rich retrospective data available on MCP to emulate the conditions of traditional randomized controlled trials (RCTs). By doing so, it enables high-quality research that sidesteps the usual costs and ethical concerns associated with traditional RCTs. More specifically, we developed methodologies to emulate RCTs for evaluating drug efficacy in HF patients using real-world observational data. Key objectives include identifying suitable RCT candidates for emulation and leveraging EHR data to replicate heart failure drug efficacy trials, thereby enabling robust comparative effectiveness research in the absence of traditional RCTs. Additionally, this project explores the use of the Cohort Visualizer, a code-free analytical tool designed for researchers without a data science background, facilitating accessible and efficient cohort analysis.

**Project 2. Impact of antihypertensive medications (AHMs) on Alzheimer's Disease and Related Dementias (ADRD) risk in hypertensive patients with mild cognitive impairment (MCI).** This study aims to validate findings from a prior study[11] that suggested AHM use may be associated with a reduced risk of ADRD in hypertensive patients with MCI. Utilizing real-world observational data, the primary objective is to perform survival analysis to assess the relationship between AHM use and ADRD progression. Additionally, the study investigates potential drug-drug interactions between AHMs, statins, and metformin within the target patient cohort, providing further insights into pharmacological influences on dementia risk. This project serves as a simulation of traditional clinical research, employing statistical analysis to assess real-world evidence.

**Project 3. Building a Mild Cognitive Impairment (MCI) to Alzheimer's Disease (AD) progression prediction model using EHR data and deep learning method.** This project focuses on training and validating a deep learning model[12] to predict the progression from MCI to AD using longitudinal EHR data. Specifically, it employs the Bidirectional Gated Recurrent Units (BiGRU) deep learning model to forecast MCI progression at varying time intervals, extending up to five years post-diagnosis. Additionally, the study aims to validate the model's generalizability across diverse datasets and healthcare systems, ensuring its applicability in real-world clinical settings.

**Project 4. Developing Deep Learning Model to predict Major Adverse Cardiovascular Events (MACE) After Liver Transplantation (LT).** This project focuses on leveraging longitudinal EHR data to develop advanced deep learning models on the MCP for predicting MACE following LT. Utilizing the MCP, we have developed innovative AI methodologies that not only enhance prediction accuracy but also outperform our previous models[13]. By

identifying high-risk candidates, the model aids clinicians in risk stratification and informs management strategies to improve transplant outcomes. Additionally, the model highlights key predictive features, enabling physicians to implement targeted preventive measures to reduce the likelihood of adverse cardiovascular events. This study demonstrates the capability of MCP in facilitating deep learning model development for clinical research.

**Data Collection and Analysis Approach**

The MCP Discover tools have played a crucial role in facilitating these projects by providing a unified platform for cohort development, data extraction, and analysis. Specifically, Project 1 leveraged the Cohort Visualizer to identify RCT candidates. Subsequently, all projects utilized Jupyter Notebook to execute SparkSQL API queries for extracting EHR data from the MCP database. Finally, data analysis—including statistical evaluations and deep learning modeling—was conducted within the Workspace using either R or Python.

**Results**

**Table 1.** Overview of Clinical Research Projects Leveraging Mayo Clinic Platform (MCP)

| Project | Project 1: RCT Emulating for HF Drug Efficacy | Project 2: AHMs Impact on ADRD | Project 3: MCI-to-AD Progression Prediction | Project 4: MACE after LT prediction |
|---|---|---|---|---|
| **Research Method** | Statistical Analysis | Statistical Analysis | Deep Learning | Deep Learning |
| **MCP Tools Involved** | Cohort Visualizer, Schema Visualizer, Workspaces | Schema Visualizer, Workspaces | Schema Visualizer, Workspaces | Schema Visualizer, Workspaces |
| **MCP Data Involved** | Structured EHR data (demographic, diagnosis, procedure, lab test, medication) | Structured EHR data (demographic, diagnosis, procedure, lab test, medication) | Structured EHR data (demographic, diagnosis, procedure, medication) | Structured EHR data (demographic, diagnosis, procedure, medication) |
| **Deliverable Outcomes by using MCP** | 1) Pipeline for emulation of drug efficacy RCTs, 2) Potential research paper | 1) Scientific validation for existing study, 2) Potential research paper | 1) AI model for MCI-to-AD prediction, 2) Potential research paper | 1) AI model for MACE after LT prediction, 2) Potential research paper |

Table 1 summarizes four clinical research projects conducted using the Mayo Clinic Platform (MCP), which includes both traditional statistical analysis and AI-based research. For the randomized controlled trial (RCT) emulation project, the Cohort Visualizer tool was used to build the study cohort. All projects also utilized the Schema Visualizer and Workspaces for EHR data collection and analysis. The results highlight the effectiveness of MCP's data, tools, and computing environment in facilitating successful data science research. MCP contributed to significant outcomes across all projects, including the development of a reusable research pipeline, scientific validation of existing studies, and AI-based prediction model. Furthermore, each project produced publication-ready results, tables, and figures for potential research papers.

**Discussion and Conclusion**

In this study, we demonstrated that MCP has played a critical role in enabling clinical studies using real-world EHR data. The MCP provides not only comprehensive, standardized, and de-identified real-world data, but also powerful tools in the data science and healthcare domains. We appreciated key features such as the Cohort Visualizer, Schema Visualizer, and Workspaces. To comprehensively showcase MCP's capabilities in clinical research, we designed four distinct projects. Project 1 explored the use of a code-free tool for clinical studies. Project 2 validated medical research through statistical analysis. Project 3 validated a deep learning model using a different dataset. Project 4 investigated MCP's ability to support deep learning model development. Our studies not only yielded publishable results from the research perspective, but also effectively leveraged AI-driven methodologies to address real-world clinical challenges, reinforcing the platform's impact on both academic research and clinical innovation.

Compared to traditional institutional EHR database, the Mayo Clinic Platform (MCP) offers distinct advantages for clinical research (as Table 2 shows). MCP provides de-identified data, streamlining IRB approvals and accelerating research timelines for users. Additionally, it enables external researchers to access high-quality Mayo EHR data for study validation and analysis, whereas institutional EHR database is usually restricted to internal use. MCP also incorporates extensive data standardization, particularly for unstructured clinical notes, by offering pre-processed, synthesized standard data. In contrast, most of the traditional institutional EHR databases primarily used standard medical terminologies[14]. Furthermore, MCP supports a broad range of users with tools ranging from code-free interfaces to advanced coding environments, making it more accessible. In comparison, using institutional EHR database is more coding-intensive, requiring a steeper learning curve. Moreover, MCP will not only integrate Mayo's data but also data from other institutions, such as Mercy, thereby broadening the scope of available research data. By offering these capabilities, MCP enhances data analysis, improves model validation, and facilitates more efficient and reliable clinical research.

**Table 2.** Advantage compares with institutional EHR database.

| Feature | MCP | Institutional EHR database |
|---|---|---|
| **Data Access** | Available for Mayo and external researchers | Restricted to internal use |
| **Data Type** | De-identified data | Identifiable data |
| **Data Standardization** | Extensive standardization | Limited standardization |
| **Tool Accessibility** | Supports both code-free and coding-dependent tools | More coding-intensive |
| **Learning Curve** | Lower, accommodates various user skill levels | Higher, due to coding dependency |
| **Data Integration** | Integrates Mayo data and data from other institutions | Internal data within the institution |

A limitation of this study is that all four projects focused exclusively on structured EHR data within MCP, without incorporating other data modalities. In the future, we plan to use additional data types of MCP, including clinical notes, medical images, and omics data, to broaden research opportunities. Furthermore, as external datasets become available, cross-validation across institutions will further strengthen clinical research. Last but not least, MCP Deploy provides a state-of-the-art infrastructure designed to streamline the integration of AI solutions into clinical workflows. While we have yet to implement these four research projects in Deploy, future studies will explore its capabilities to facilitate AI deployment and assess its potential to accelerate the translation of AI-driven innovations into real-world clinical practice. By leveraging MCP, we aim to bridge the gap between research and clinical application.

In the era of AI, the MCP is poised to revolutionize clinical research by advancing multimodal AI, real-world evidence generation, and global data collaboration. By integrating structured EHR data, clinical notes, imaging, and genomics, researchers can leverage MCP harmonized data to enhance biomedical knowledge for large medical foundation models. This integration will boost downstream tasks such as predictive analytics for early disease detection and personalized treatment[3-4]. MCP also ensure robust and generalizable AI model validation across multiple institutions. Its data ecosystem will facilitate large-scale studies while maintaining patient privacy, effectively bridging the gap between AI research and real-world clinical implementation[15]. Additionally, MCP will transform drug development by enabling real-world evidence-based trials that extend beyond traditional clinical settings. This approach allows for broader participation and more diverse data collection, enhancing trial efficiency and relevance[16]. In addition, MCP can facilitate our "Clinical Trials Beyond Walls" approach which allows broader participation by removing barriers for patient involvement and includes initiatives with underserved communities to enhance the relevance and quality of clinical trials[17]. With scalable research tools and expanded accessibility, MCP will empower a diverse research community, accelerating medical innovation and driving the future of precision medicine and proactive healthcare.

**Data availability**

This study involves analysis of de-identified Electronic Health Record (EHR) data via Mayo Clinic Platform Discover. Data shown and reported in this manuscript has been extracted from the EHR using an established protocol for data extraction, aimed at preserving patient privacy. The data has been determined to be de-identified pursuant to an expert's evaluation, in accordance with the HIPAA Privacy Rule. Any data beyond what is reported in the manuscript, including but not limited to the raw EHR data, cannot be shared or released due to the parameters of the expert determination to maintain the data de-identification. Contact corresponding authors for additional details regarding Mayo Clinic Platform Discover.